\documentclass[12pt]{article}

\usepackage{graphicx}

\input{epsf.sty}

\def\fo{\hbox{{1}\kern-.25em\hbox{l}}}

\def\slashchar#1{\setbox0=\hbox{$#1$}           
   \dimen0=\wd0                                 
   \setbox1=\hbox{/} \dimen1=\wd1               
   \ifdim\dimen0>\dimen1                        
      \rlap{\hbox to \dimen0{\hfil/\hfil}}      
      #1                                        
   \else                                        
      \rlap{\hbox to \dimen1{\hfil$#1$\hfil}}   
      /                                         
   \fi}                                         %

\def\hide#1{[hidden stuff]}

\def\beq{\begin{equation}}
\def\eeq{\end{equation}}
\def\eq{\end{equation}}
\def\to{\rightarrow}

\def\mEt{\mbox{${\hbox{$E$\kern-0.6em\lower-.1ex\hbox{/}}}_T$}\, } 

\def\bsg{\ifmmode B\to X_s\gamma\else $B\to X_s\gamma$\fi}
\def\bsll{\ifmmode B\to X_s\ell^+\ell^-\else $B\to X_s\ell^+\ell^-$\fi}
\def\bstt{\ifmmode B\to X_s\tau^+\tau^-\else $B\to X_s\tau^+\tau^-$\fi}
\def\shat{\ifmmode \hat{s}\else $\hat{s}$\fi}

\newcommand{\newc}{\newcommand}

\newc{\asusy}{\delta a^{\rm SUSY}_\mu}
\newc{\lcal}{\int {\cal L}dt}
 
\newc{\LSP}{{\chi^0_1}}
\newc{\stauR}{{\tilde \tau_R}}
\newc{\stau}{{\tilde \tau_1}}
\newc{\mstop}{m_{\tilde{t}}}
\newc{\mHpm}{m_{H^\pm}}
\newc{\gsim}{\lower.7ex\hbox{$\;\stackrel{\textstyle>}{\sim}\;$}}
\newc{\lsim}{\lower.7ex\hbox{$\;\stackrel{\textstyle<}{\sim}\;$}}
\newc{\ie}{{\it i.e.}}          
\newc{\etal}{{\it et al.}}
\newc{\eg}{{\it e.g.}}          
\newc{\kev}{\hbox{\rm\,keV}}            
\newc{\mev}{\hbox{\rm\,MeV}}            
\newc{\gev}{\hbox{\rm\,GeV}}            
\newc{\tev}{\hbox{\rm\,TeV}}
\newc{\xpb}{\hbox{\rm\, pb}}
\newc{\xfb}{\hbox{\rm\, fb}}

\newc{\mtop}{m_t}
\newc{\mbot}{m_b}
\newc{\mz}{m_Z}
\newc{\mw}{M_W}
\newc{\alphasmz}{\alpha_s(m_Z^2)}
\newc{\swsq}{\sin^2\theta_W}
\newc{\tw}{\tan\theta_W}
\newc{\cw}{\cos\theta_W}
\newc{\sw}{\sin\theta_W}
\newc{\BR}{\hbox{\rm BR}}
\newc{\zbb}{Z\to b\bar}
\newc{\Gb}{\Gamma (Z\to b\bar b)}
\newc{\Gh}{\Gamma (Z\to \hbox{\rm hadrons})}
\newc{\rbsm}{R_b^\hbox{\rm sm}}
\newc{\rbsusy}{R_b^\hbox{\rm susy}}
\newc{\drb}{\delta R_b}

\newc{\sgn}{\mbox{sgn}}
\newc{\tbeta}{\tan\beta}
\newc{\uL}{{\tilde u_L}}
\newc{\uR}{{\tilde u_R}}
\newc{\cL}{{\tilde c_L}}
\newc{\cR}{{\tilde c_R}}
\newc{\tL}{{\tilde t_L}}
\newc{\tR}{{\tilde t_R}}
\newc{\dL}{{\tilde d_L}}
\newc{\dR}{{\tilde d_R}}
\newc{\sL}{{\tilde s_L}}
\newc{\sR}{{\tilde s_R}}
\newc{\bL}{{\tilde b_L}}
\newc{\bR}{{\tilde b_R}}
\newc{\eL}{{\tilde e_L}}
\newc{\eR}{{\tilde e_R}}
\newc{\mhp}{m_{H^\pm}}
\newc{\mhalf}{m_{1/2}}
\newc{\emt}{{e/\mu /\tau}}

\newc{\lR}{\tilde{l}_R}
\newc{\lL}{\tilde{l}_L}
\newc{\nL}{\tilde{\nu}_L}
\newc{\na}{\chi^0_1}
\newc{\nb}{\chi^0_2}
\newc{\nc}{\chi^0_3}
\newc{\nd}{\chi^0_4}
\newc{\ca}{\chi^{\pm}_1}
\newc{\cb}{\chi^{\pm}_2}
\newc{\camp}{\chi^\mp_1}
\newc{\cbmp}{\chi^\mp_1}
\newc{\capos}{\chi^{+}_1}
\newc{\caneg}{\chi^{-}_1}
\newc{\phit}{\phi_t}
\newc{\phib}{\varphi_b}
\newc{\phiew}{\phi_{ew}}
\newc{\htz}{h^0_t}
\newc{\hbz}{h^0_b}
\newc{\hewz}{h^0_{ew}}
\newc{\hsmz}{h^0_{sm}}
\newc{\huz}{h^0_u}
\newc{\hsusyz}{h^0_{susy}}

\newcommand{\drawsquare}[2]{\hbox{%
\rule{#2pt}{#1pt}\hskip-#2pt
\rule{#1pt}{#2pt}\hskip-#1pt
\rule[#1pt]{#1pt}{#2pt}}\rule[#1pt]{#2pt}{#2pt}\hskip-#2pt
\rule{#2pt}{#1pt}}

\newc{\Dal}{\drawsquare{7}{0.6}}

\def\beq{\begin{equation}}
\def\eeq{\end{equation}}
\def\bea{\begin{eqnarray}}
\def\eea{\end{eqnarray}}
\catcode`@=11
\long\def\@caption#1[#2]#3{\par\addcontentsline{\csname
  ext@#1\endcsname}{#1}{\protect\numberline{\csname
  the#1\endcsname}{\ignorespaces #2}}\begingroup
    \small
    \@parboxrestore
    \@makecaption{\csname fnum@#1\endcsname}{\ignorespaces #3}\par
  \endgroup}
\catcode`@=12

\begin{document}
\begin{titlepage}

\begin{flushright}
hep-ph/0202075 \\
LBNL-49595\\
UCD 2001-14
\end{flushright}

\huge
\bigskip
\bigskip
\begin{center}
{\Large\bf
Proton Decay due to bulk SU(5) gauge bosons in the Randall-Sundrum Scenario}
\end{center}

\large

\vspace{.15in}
\begin{center}

Shrihari Gopalakrishna

\small

\vspace{.1in}
{\it Physics Department, 
               University of California, Davis, CA 95616}, and\\
{\it Theoretical Physics Group, Lawrence Berkeley National 
 Laboratory, Berkeley, CA 94720}\\

\end{center}

\vspace{0.15in}
 
\begin{abstract}
We calculate the proton decay rate due to five dimensional SU(5) gauge bosons in the Randall-Sundrum scenario with two branes (Planck brane and TeV brane). We consider matter in the usual $10$ and $\bar 5$ SU(5) representations localized on a brane, and consider the case when SU(5) is broken by a Higgs mechanism on the matter brane. We calculate the proton decay rate mediated by $X$ bosons, due to terms in the Lagrangian with $X_\mu$ and its derivative along the extra dimension. We confirm that the experimental limit on the proton decay rate allows SU(5) matter to be on the Planck brane, but excludes it from being on the TeV brane in this scenario.
\end{abstract}

\medskip

\begin{flushleft}
February 2002
\end{flushleft}

\end{titlepage}

\baselineskip=18pt

\section{Introduction}

There is a hierarchy of some 17 orders of magnitude between the Planck scale ($M_{pl} \sim 10^{19}\gev$) and the Electro-weak scale ($M_{EW} \sim 10^2\gev$). The Standard Model (SM) requires an extreme fine tuning to maintain this hierarchy and a lot of recent work has addressed this ``Hierarchy Problem''. Supersymmetry, Technicolor and recently, extra dimensions~\cite{Arkani-Hamed:1998rs}\,\cite{Randall:1999ee}, have been considered to address this problem.

One of the frameworks that addresses the hierarchy problem was proposed by Randall and Sundrum~\cite{Randall:1999ee}. They propose a 5-dimensional non-factorizable AdS geometry with the metric given by:
\beq
ds^2 = g_{MN}~dx^M dx^N = g_{\mu\nu}~dx^\mu dx^\nu + dy^2 = e^{-2k|y|}\eta_{\alpha\beta}~dx^\alpha dx^\beta + dy^2
\eeq
where, $k$ is the AdS curvature, $y$ is the co-ordinate in the extra dimension and $\eta_{\alpha\beta}$ is $diag(-1,1,1,1)$.

The extra dimension is compactified on $S^1/Z_2$. The $Z_2$ symmetry identifies $y$ and $-y$ and fields are either even or odd with respect to this $Z_2$: 
\beq
\Phi (x, -y) = \pm \Phi (x,y).
\eeq
The Lagrangian is constructed to be be $Z_2$ invariant. In this setup there are two 3-branes at the $Z_2$ orbifold fixed points~-~the ``Planck Brane'' at $y=0$, and the ``TeV Brane'' at $y=\pi R$. The ``warp factor'', $e^{-2k|y|}$, allows the possibility of generating an exponentially lower Electro-Weak scale, given a high Planck scale; $M_{EW} \approx e^{-k\pi R}M_{pl}$, with $R$ the radius of the compact extra dimension. To generate the required hierarchy we need $k R \approx 12$ with $k \sim M_{pl}$. In the original framework, to solve the hierarchy problem, matter was thought to reside on the TeV brane; later authors have considered matter on the Planck brane also. A generic feature of Randall-Sundrum type theories is that it requires the cosmological constants in the bulk and the two branes to be fine-tuned\footnote{Ref.~\cite{Goldberger:1999uk} describes a mechanism to stabilize the radius of the compact extra dimension and reduce the number of fine-tunings to one.}. 
\bea
\Lambda = -6 M_5^3 k^2 \nonumber \\ 
\Lambda_{(0)} = - \Lambda_{(\pi R)} = - \frac{\Lambda}{k}
\eea
where $M_5$ is the fundamental five-dimensional scale. The four dimensional Planck scale is:
\beq
M_{pl}^2 = \frac{M_5^3}{k}(1-e^{-2\pi k R}).
\eeq
In this paper, we work in this framework of Randall and Sundrum.

Grand Unified Theories (GUT) based on $SU(5)$ have many compelling features\cite{Hall:TASI}. The quantum numbers of particles in the SM can be neatly explained by considering them to be in the $10$ and $\bar 5$ representations. Another motivation to consider GUTs is the apparent unification of gauge couplings at around $M_{GUT} \sim 10^{16}\gev$, in the Minimal Supersymmetric Standard Model (MSSM)~\cite{Dimopoulos:1981yj}.

Our motivation for considering extra dimensional gauge fields is the possibility of bulk gauge fields mediating supersymmetry breaking through Gaugino Mediation~\cite{Kaplan:1999ac}. In this scheme, a gauge supermultiplet in the bulk of a flat higher dimensional theory transmits supersymmetry breaking at a TeV from the source brane to the MSSM fields on a second brane. The realization of Gaugino Mediation in the context of the Randall-Sundrum scenario is left for future work.

Motivated by these considerations we consider a setup with $SU(5)$ gauge bosons in the bulk of the Randall-Sundrum setup. We consider matter in the $10$ and $\bar 5$ representations of $SU(5)$ to be localized on a brane. Pomarol~\cite{Pomarol:2000hp} has considered breaking $SU(5)$ by a Higgs mechanism on the matter brane in the Randall-Sundrum scenario. In this paper we calculate the proton decay rate due to terms with an $X$ boson and also due to non-renormalizable interaction terms containing $\partial_y X$.

%
%

\section{The SU(5) Lagrangian}
In the following, only terms relevant for proton decay will be shown. The $SU(5)$ generators are denoted as $T^a$, normalized as $\{T^a, T^b\}=\frac{1}{2}~\delta^{ab}$ and $y_0$ is the position of the matter brane.\\
The field strength for the bulk gauge field $A^a_M$ (of mass dimension $3/2$) is defined as:\\
\mbox{    } $F_{MN} = \partial _M A_N - \partial _N A_M - ig\left[A_M, A_N \right] $\\
\mbox{    } $F_{MN} = F^a_{MN}T^a; A_M = A^a_M T^a, etc.$\\
and the covariant derivatives for the usual GUT matter multiplets, $T$ the $10$ of $SU(5)$ and $\bar F$ the $\bar 5$ of $SU(5)$ is given by:\\
\mbox{    } $D_\mu T = \partial_\mu T - i \frac{e}{\sqrt{2 M_{pl}}} (A_\mu T + T A_\mu^T)$\\
\mbox{    } $D_\mu F = \partial_\mu F - i \frac{e}{\sqrt{2 M_{pl}}} A_\mu F$.

The $SU(5)$ invariant lagrangian is: 
\bea
L &=& L_{Bulk}~+~L_{Brane} \\
L_{Bulk} &=& \sqrt{-g}\left[ \frac{1}{2}~Tr(F^{MN}F_{MN}) \right] \\
L_{Brane} &\supset& \sqrt{-g}~\delta(y-y_0) \left[ Tr(\bar T i\slashchar{D} T) + \frac{i g_1}{M_{CO}\sqrt{M_{pl}}} Tr\left( \bar T \left[ \gamma^M, \gamma^N \right] F_{MN} T \right) \right. \nonumber \\
&&\left. +~\bar F i \slashchar{D} F + \frac{i g_2}{M_{CO}\sqrt{M_{pl}}}\left( \bar F \left[ \gamma^M, \gamma^N \right] F_{MN}\,F\right) \right]
\label{eqLBR}
\eea
where, $g$ is the determinant of the 5-dimensional metric \footnote{Note: $\sqrt{-g} \sim e^{-4k|y|}$}, $\gamma_M = e_M^\alpha \gamma_{\alpha}$ ($e_M^\alpha$ is the vierbein and $\gamma_{\alpha}$ are the flat space Dirac matrices) and the second and fourth terms above are non-renormalizable interaction terms that are Lorentz and $SU(5)$ invariant, suppressed by the cutoff scale $M_{CO}$, and $e$, $g_1$ and $g_2$ are $O(1)$ coupling constants . 

For the Lagrangian to be $Z_2$ invariant, $A^a_M$ should have the property: 
\bea
A^a_\mu (x, -y) = + A^a_\mu (x,y) 
\label{eqAmuZ2par}\\
A^a_5 (x, -y) = - A^a_5 (x,y)
\label{eqA5Z2par}
\eea
i.e., $A^a_\mu$ is $Z_2$ even and $A^a_5$ is $Z_2$ odd.

The equations of motion for the $A^a_M$ are:
\bea
\partial_\alpha\partial^\alpha A^a_5 &-& \partial_y\partial_\alpha A^{a\alpha} = 0 
\label{eqAEQMOT1F} \\
e^{2k|y|} \partial_y\left(e^{-2k|y|} \left(\partial_y A^{a\beta} - \partial^\beta A^a_5 \right) \right) &+& e^{2k|y|} \partial_\alpha \left(\partial^\alpha A^{a\beta} - \partial^\beta A^{a\alpha} \right) = 0.
\label{eqAEQMOT2F}
\eea
Using the gauge freedom, we can always set $\partial_\mu A^{a\mu} = 0$. Then the equation of motion eq.\,(\ref{eqAEQMOT1F}) gives $A^a_5 = 0$.\footnote{Though $A^a_5$ is not identically zero off-shell, it will not contribute to proton decay since it is $Z_2$ odd and therefore vanishes at the fixed points where $SU(5)$ matter is located.} In this on-shell gauge eq.\,(\ref{eqAEQMOT2F}) becomes:
\beq
e^{2k|y|} \partial_y\left(e^{-2k|y|}\partial_y A^{a\beta} \right) + e^{2k|y|} \partial_\alpha\partial^\alpha A^{a\beta} = 0.
\label{eqAEQMOT}
\eeq

It is helpful to expand into Kaluza-Klein modes\footnote{Sum over $n$ is understood.}:
\beq
A_M(x,y) = \frac{1}{\sqrt{2\pi R}} A_M^{(n)}(x) f_n(y)
\label{eqAKK}
\eeq
with the $f_n$ a complete set of functions over $y$ that satisfy:
\beq
\frac{1}{2\pi R} \mathop{\int}_{-\pi R}^{\pi R} dy ~f_mf_n ~=~ \delta_{mn}.
\eeq
Eq.\,(\ref{eqAEQMOT}) can then be split into the two equations:
\bea
\partial_\alpha\partial^\alpha A^{(n)\beta}(x) = m_n^2 A^{(n)\beta}(x) 
\label{eqAKK4D}\\
e^{2k|y|} \partial_y\left(e^{-2k|y|}\partial_y f_n(y) \right) + e^{2k|y|} m_n^2 f_n(y) = 0 
\label{eqFNY}
\eea
where $m_n$ is a discrete sequence (KK masses)\footnote{The $m_n$ will be shown to have a gap $\sim\tev$.} determined by the boundary (brane) conditions on the $f_n$ due to its $Z_2$ property.
Eq.\,(\ref{eqFNY}) can be reduced to the Bessels Differential Equation and the solutions~\cite{Goldberger:1999wh}\,\cite{Gherghetta:2000qt} (with definite $Z_2$ even and odd properties, $f_n^{even}$ and $f_n^{odd}$) contain the Bessels functions, $J_1$ and $Y_1$. 
\bea
f_n^{even}(y) &=& \frac{e^{k|y|}}{N_n}\left[ J_1\left( \frac{m_n}{k} e^{k|y|} \right)~+~ b_1(m_n)~Y_1\left( \frac{m_n}{k} e^{k|y|} \right) \right]
\label{eqFNEVN} \\
f_n^{odd}(y) &=& \theta(y)~f_n^{even}(y) 
\label{eqFNODD} \\
N_n^2 &=& \frac{1}{\pi R} \mathop{\int}_{0}^{\pi R} dy~e^{2ky}\left[ J_1\left( \frac{m_n}{k} e^{ky} \right)~+~ b_1(m_n)~Y_1\left( \frac{m_n}{k} e^{ky} \right) \right] ^2
\label{eqFNNORM}
\eea
where $\theta(y)$ is the step function and $b_1(m_n)$ is a constant that depends on $m_n$. In the limit $m_n \ll k$ and $kR \gg 1$, $N_n$ is approximated by~\cite{Gherghetta:2000qt}:  
\beq
N_n \approx \frac{e^{\pi k R}}{\sqrt{2\pi k R}} J_1(\frac{m_n}{k} e^{\pi k R}).
\eeq
Using the large argument approximation for the Bessels function from eq.\,(\ref{eqJ1LARGE})
\beq
N_n \approx \frac{e^{\frac{\pi k R}{2}}}{\pi \sqrt{R m_n}}.
\label{eqAPRNORM}
\eeq

Since we are interested in analyzing the proton decay rate, let us focus on the terms involving the $T$ and $X$ bosons in eq.\,(\ref{eqLBR})\footnote{The terms with $\bar F$ lead to similar contributions as the $T$ and will not be explicitly considered here.}: 
\beq
L_{Brane} \supset \sqrt{-g}~\delta(y-y_0)\left[ \left( \frac{-ie}{\sqrt{2 M_{pl}}}\right) Tr\left(\bar T i \gamma^M A_M T\right) + \frac{i g_1}{M_{CO}\sqrt{M_{pl}}} Tr\left( \bar T \left[ \gamma^M, \gamma^N \right] F_{MN} T \right)\right].
\label{eqLTTA}
\eeq
The first term above is contained in the first term in eq.\,(\ref{eqLBR}). The $X_M$ part of $A_M$ can lead to proton decay due to $uu\rightarrow e^+ \bar d$.

Using eq.\,(\ref{eqAKK}) in eq.\,(\ref{eqLTTA}), integrating over $y$ (which picks out the fixed point values) and bringing to canonical form with $T \rightarrow T~e^{\frac{3}{2}k|y_0|}$, we get the effective four dimensional lagrangian:
\beq
L^{(4)} \supset \frac{-ie}{\sqrt{2 M_{pl}}} \frac{f_n(y_0)}{\sqrt{2 \pi R}} Tr(\bar T i \gamma^\alpha A_\alpha^{(n)} T) + \frac{2 i g_1}{M_{CO}\sqrt{M_{pl}}} \frac{(\partial_y f_n)|_{y_0}}{\sqrt{2\pi R}} Tr\left(\bar T \left[ \gamma^\alpha, \gamma^5 \right] A_\alpha^{(n)} T\right).
\label{eqTATC}
\eeq

\section{SU(5) breaking by the Higgs mechanism}
We do not see the $SU(5)$ structure at the weak scale and so it has to be broken at a higher scale. There are various ways of breaking $SU(5)$ and one of them is by a Higgs confined to the matter brane. Let us consider a Higgs in the adjoint of $SU(5)$ localized on the matter brane~\cite{Pomarol:2000hp} and that it gets a VEV, breaking $SU(5)$ down to $SU(3)\times SU(2)\times U(1)$ in the usual manner~\cite{Hall:TASI}. 

The Higgs mechanism results in a mass term for the $X$ and $Y$ bosons which effectively can be parametrized as:
\beq
L_{Brane} \supset \sqrt{-g}~\delta(y-y_0)~\frac{v^2}{M_{pl}}~A^{\hat a}_M {A^{\hat a}}^M
\eeq
where, $v$ is the VEV of the Higgs that breaks $SU(5)$ to $SU(3)\times SU(2)\times U(1)$. We will not show terms containing the physical Higgs since we are interested in calculating the proton decay contribution due to the gauge bosons. We will work in the Unitary gauge where we only deal with physical particles.

The equation of motion for gauge bosons (eq.\,(\ref{eqAEQMOT})), in the presence of such a brane mass term is modified to:
\beq
e^{2k|y|}\partial_y\left(e^{-2k|y|}\partial_y A^{a\hat\beta} \right) + e^{2k|y|} \partial_\alpha\partial^\alpha A^{a\hat\beta} = 2\delta(y-y_0)\frac{v^2}{M_{pl}} A^{a\hat\beta}.
\label{eqAMEQMOT}
\eeq
Eq.\,(\ref{eqAKK4D}) remains unchanged, but eq.\,(\ref{eqFNY}) is modified to:
\beq
e^{2k|y|} \partial_y\left(e^{-2k|y|}\partial_y f_n(y) \right) + e^{2k|y|} m_n^2 f_n(y) = 2\delta(y-y_0)\frac{v^2}{M_{pl}} f_n(y).
\label{eqMFNY}
\eeq

Since the the $A_\mu$ are $Z_2$ even (see eq.\,(\ref{eqAmuZ2par})), we require the corresponding $f_n$ to satisfy ($y_0$ is the matter brane and $y_0'$ the other brane):
\bea
\frac{d}{dy}f_n|_{y_0} &=& \frac{2v^2}{M_{pl}} f_n(y_0)
\label{eqJMP1} \\
\frac{d}{dy}f_n|_{y_0'} &=& 0.
\label{eqJMP2}
\eea 
Eq.\,(\ref{eqJMP1}) implies for $b_1(m_n)$ (the constant defined in eq.\,(\ref{eqFNEVN})):
\beq
b_1(m_n) = -\frac{ (1-\frac{v^2}{k M_{pl}}) J_1(\frac{m_n}{k}e^{k y_0}) + \frac{m_n}{k}e^{k y_0} J_1'(\frac{m_n}{k}e^{k y_0})} { (1-\frac{v^2}{k M_{pl}}) Y_1(\frac{m_n}{k}e^{k y_0}) + \frac{m_n}{k}e^{k y_0} Y_1'(\frac{m_n}{k}e^{k y_0})}
\label{eqb1y0}
\eeq
and Eq.\,(\ref{eqJMP2}) implies:
\beq
b_1(m_n) = -\frac{J_1(\frac{m_n}{k}e^{k y_0'}) + \frac{m_n}{k}e^{k y_0'} J_1'(\frac{m_n}{k}e^{k y_0'})} {Y_1(\frac{m_n}{k}e^{k y_0'}) + \frac{m_n}{k}e^{k y_0'} Y_1'(\frac{m_n}{k}e^{k y_0'})}.
\label{eqb1y0pri}
\eeq
Equating these determines $m_n$ as we show next. 

\subsection{Matter on the Planck brane.}
For matter on the Planck brane, $y_0 = 0$ and $y_0' = \pi R$. The matter content on the Planck Brane is that of the MSSM and the gauge couplings apparently unify~\cite{Pomarol:2000hp}\,\cite{Randall:2001gc} in the usual manner\footnote{The authors of ref.~\cite{Arkani-Hamed:2000ds} point out that there may be large threshold corrections that upset gauge coupling unification, but there is still the possibility that the threshold corrections are universal and preserve unification.}, at around $M_{GUT}\sim 10^{16}\gev$. 

Equating eqs.~(\ref{eqb1y0}) and (\ref{eqb1y0pri}) determines $m_n$. Using eq.\,(\ref{eqJYREC}) this equality becomes:
\beq
\frac{-\frac{v^2}{k M_{pl}} J_1(\frac{m_n}{k}) + \frac{m_n}{k} J_0(\frac{m_n}{k})}{-\frac{v^2}{k M_{pl}} Y_1(\frac{m_n}{k}) + \frac{m_n}{k} Y_0(\frac{m_n}{k})} = \frac{J_0\left(\frac{m_n}{k}e^{k\pi R}\right)}{Y_0\left(\frac{m_n}{k}e^{k\pi R}\right)}.
\eeq
Using eq.\,(\ref{eqJYSMALL}) for $\frac{m_n}{k} \ll 1$ on the left hand side, and defining $u_n = \frac{m_n}{k}e^{k \pi R}$ the above equation becomes:
\beq
\frac{\pi}{2}\frac{1}{\ln\left(\frac{u_n}{2}\right) - k\pi R + \gamma + \frac{v^2}{k M_{pl}}\frac{e^{2k\pi R}}{u_n^2}} = \frac{J_0\left(u_n\right)}{Y_0\left(u_n\right)}
\label{eqb1lhssm}
\eeq 
where $\gamma \approx 0.5772$.

With no Higgs VEV, i.e. $v=0$, the lowest lying mode is the massless zero mode obtained as the limit $u_n \rightarrow 0$ in the above equation. The corresponding limit for the wavefunction from eq.\,(\ref{eqFNEVN}) is $f_n(y) = {\rm constant}$.

Now we ask whether this ``zero-mode'' becomes massive due to a non-zero $v$. For $u \ll 1$, the right hand side of eq.\,(\ref{eqb1lhssm}) can be approximated using eq.\,(\ref{eqJYSMALL}):
\beq
\frac{\pi}{2}\frac{1}{\ln\left(\frac{u_n}{2}\right) - k\pi R + \gamma + \frac{v^2}{k M_{pl}}\frac{e^{2k\pi R}}{u_n^2}} = \frac{\pi}{2}\frac{1}{\ln\left(\frac{u_n}{2}\right) + \gamma}.
\eeq
The lowest mode (what was massless in the $v=0$ case) that solves this is given by:
\beq
u_0 = \frac{v}{\sqrt{k\pi R}} \frac{1}{\sqrt{k M_{pl}}} e^{k\pi R},
\eeq
or equivalently,
\beq
m_0 = \frac{v}{\sqrt{\pi M_{pl}R}}.
\label{eqMVSMALL}
\eeq
It should be stressed that this solution for the lowest mode is valid in the limit $u \ll 1$, i.e., $v \ll 10^3\gev$. Thus the mode that was massless is now lifted to the value given by eq.\,(\ref{eqMVSMALL}) due to a non-zero $v$. We are actually interested in the case when $v \sim M_{GUT}$ for which we cannot use the above solution obtained for $u \ll 1$. A numerical computation of eq.\,(\ref{eqb1lhssm}) shows that as $v$ becomes larger than about $10^4\gev$ the lowest mode gets fixed at around the first zero of $Y_0$, $u_0 \approx 0.9$, or equivalently $m_0 \approx 0.9 k e^{-k\pi R} \sim\tev$. Thus we see that the zero mode is lifted to a mass of $\sim\tev$ due to a VEV $v \sim M_{GUT}$. The wavefunction of this ``zero-mode'' is similar to the wavefunction of the first KK mode since their masses are close. 

We can see from eq.\,(\ref{eqb1lhssm}) that the low lying KK modes, $u_n$, are close to the zero's of the Bessel function $J_0$, namely, $u_n \approx \left(n-\frac{1}{4}\right)\pi$ which translates to:
\beq
m_n \approx \left(n-\frac{1}{4}\right)\pi k e^{-k \pi R}.
\label{eqMN}
\eeq
Thus, a non-zero $v$ doesn't significantly change the mass spectrum from the $v=0$ case (see for example ref.~\cite{Gherghetta:2000qt}). We will estimate the wavefunctions at $y=0$ in the next section and we will see that it changes drastically when a VEV is turned on.

\subsection{Matter on the TeV brane}
We will show in the next section that having matter on the TeV brane violates the experimental limit on the proton decay. Phenomenological disadvantages of having matter on the TeV brane with gauge fields in the bulk are discussed in ref.~\cite{Davoudiasl:1999tf}.

\section{Proton Decay due to Gauge Bosons}
\label{secProtDec}
The dominant contribution to proton decay comes from tree level exchange of the X and Y bosons such as $uu\rightarrow X_\mu \rightarrow e^+ \bar d$. It should be noted that $X_5$ and $Y_5$ do not induce proton decay, since, being $Z_2$ odd, they vanish at the fixed points (where matter is localized).

The proton decay rate due to the tree level exchange of the $n$th KK mode of $X_\mu$ given by the first term in eq.\,(\ref{eqTATC}) is:
\beq
\Gamma^{(n)} \approx \frac{\left(e f_n(y_0)\right)^4}{4 \left(2\pi M_{pl}R \right)^2 } \frac{m_p^5}{M_X^{(n)4}}
\label{eqGAMPORB}
\eeq
where, $m_p$ is the Proton mass, $\sim 1\gev$. The tree level contribution due to the $n$th KK mode of $\partial_y X_\mu$ given by the second term in eq.\,(\ref{eqTATC}) is:
\beq
\Gamma^{(n)} \sim \frac{\left(2 g_1\,\partial_y f_n(y_0)\right)^4}{\left(2\pi M_{pl}R\right)^2 \left(M_{CO}\right)^4} \frac{m_p^5}{M_X^{(n)4}}.
\label{eqGAMDYFN}
\eeq

Summing over the KK modes gives the total decay rate,
\beq
\Gamma = \frac{1}{\tau_p} \sim \mathop{\sum}_{n=1}^{\frac{v}{1\,TeV}} \Gamma^{(n)}.
\label{eqGAMSUMV} \\
\eeq
The sum is over the modes that are of mass less than $v$ and since the spacing of modes is approximately a TeV, we have cut the sum off at $\frac{v}{1\,TeV}$ above.

If matter is on the TeV brane, eq.\,(\ref{eqFNEVN}) implies that $f_n(\pi R) \sim O(1)$ with $m_n \sim n\tev$. Eq.\,(\ref{eqGAMPORB}) implies $\Gamma^{(1)} \sim 10^{-12}\gev$, i.e. $\tau_p^{(1)} \sim 10^{-20}$~yr, and the experimental bound on the proton decay rate\footnote{The experimental limit on the proton lifetime is (mode dependent) $\tau_p > 10^{31} - 5\times 10^{32}{\rm\,years}$~\cite{Barnett:1996hr} which translates to $\tau_p \sim 10^{64}{\rm\,(GeV)^{-1}}$.} is clearly violated. Thus, in agreement with ref.~\cite{Pomarol:2000hp} we conclude that we cannot have matter on the TeV brane. 

For matter on the Planck brane, we estimate the value of the wavefunction and its derivative along the extra dimension at $y=0$. From eqs.~(\ref{eqb1y0}) and (\ref{eqFNEVN}) with $\frac{m_n}{k} \ll 1$\footnote{From eq.\,(\ref{eqMN}) this means that all these estimates are valid for $n \ll 10^{16}$ and not for arbitrarily large $n$.}, $v>m_n$, $y_0=0$ and $N_n$ from eq.\,(\ref{eqAPRNORM}),
\bea
b_1(m_n) &\approx& - \frac{\pi^3}{2}\left(n-\frac{1}{4} \right)^2 \left(\frac{k M_{pl}}{v^2}\right) e^{-2\pi k R} \\
f_n(0) &\sim& e^{-2\pi k R} \frac{\pi^2}{\sqrt{2}} \sqrt{2\pi k R} \left(\frac{k M_{pl}}{v^2}\right) \left(n-\frac{1}{4}\right)^{\frac{3}{2}} \\
\partial_y f_n(0) &\sim&  \sqrt{2} e^{-2\pi k R} \pi k\sqrt{2\pi k R} \left(n-\frac{1}{4}\right)^{\frac{3}{2}}.
\label{eqDYFN0}
\eea

Eq.\,(\ref{eqGAMPORB}) gives the proton decay rate due to the $n$th KK mode: $\Gamma^{(n)} \sim 10^{-114}n^2$ and thus the total decay rate given by eq.\,(\ref{eqGAMSUMV}) due to the first term in eq.\,(\ref{eqTATC}) is $\Gamma \sim 10^{-75}\gev$, i.e. $\tau_p \sim 10^{43}$~yr.

We calculate next the contribution due to the term with the derivative on the X boson along the extra dimension given by the second term in eq.\,(\ref{eqTATC}).  Using eq.\,(\ref{eqDYFN0}) in eq.\,(\ref{eqGAMDYFN}) gives $\Gamma^{(n)} \sim \frac{10^{-61} n^2}{M_{CO}^4}$. With $M_{CO} \sim 10^{16}\gev$ and summing over the KK modes, the total rate due to the second term in eq.\,(\ref{eqTATC}) is: $\Gamma \sim 10^{-86}\gev$ i.e. $\tau_p \sim 10^{54}$~yr.

Thus the lifetime is far greater than the experimental limit and we conclude that it is acceptable to have matter on the Planck brane. 

\section{Conclusions}
We considered a setup based on the Randall-Sundrum scenario which addresses the hierarchy problem. This consists of having a warped geometry compactified on $S^1/Z_2$, with the MSSM matter fields localized on one of the fixed points of the orbifold. We considered the implications to proton decay due to $SU(5)$ gauge bosons in the bulk.

We considered the situation when $SU(5)$ is broken by an adjoint Higgs on the matter brane breaking it down to $SU(3)\times SU(2)\times U(1)$.  We calculated the proton decay rate due to exchange of $X$ bosons and considered the effect of a non-renormalizable interaction term on the brane having a derivative in the $y$ direction. Another way of breaking $SU(5)$ is by Orbifold breaking which was not considered here. It is possible that the contributions to proton decay can be different in Orbifold breaking due to the different $Z_2$ parity properties of the gauge bosons, although we expect the results to be qualitatively similar.      

We found that the experimental limit on the proton decay rate does not allow matter to be on the TeV brane. We then calculated the proton decay rate if the $10$ and $\bar 5$ matter multiplets of $SU(5)$ was on the Planck brane. We showed that this does not violate the experimental limit and thus is a viable setup. 

Having matter on the Planck brane implies the hierarchy problem and thus requires Supersymmetry to stabilize the Higgs at the Weak scale. Though it might seem that we lost the original motivation of solving the hierarchy problem, we note here that there is still the attractive feature of generating the Weak scale due to the warped geometry, given the Planck scale. One realization of this is that some dynamics on the TeV brane might break supersymmetry at the TeV scale and then be communicated by the Gauginos in the bulk to the MSSM on the Planck brane. We have not considered the Higgs and Higgsino contributions to Proton Decay here. It would also be interesting to consider if it is possible to have the Higgs in the bulk while satisfying proton decay constraints.

\noindent
{\it Acknowledgments:}
I would like to thank J.\,Wells (who suggested this area of work) and W.\,Goldberger for valuable discussions. I would also like to thank Z.\,Chacko, E.\,Dudas, M.\,Gaillard, J.\,Giedt, S.\,Mouslopoulos, Y.\,Nomura, M.\,Perelstein and A.\,Pomarol for their comments. I am grateful to Lawrence Livermore National Laboratory for financial support. This work was supported in part by the Director, Office of Science, Office of High Energy and Nuclear Physics, of the U.S. Department of Energy under Contract DE-AC03-76SF00098.

\appendix
\section*{Appendix}
For the readers convenience, we collect here some relations involving Bessels functions that we have used.
\bea
J_1'(x) &=& J_0(x) - \frac{1}{x} J_1(x) \nonumber \\
Y_1'(x) &=& Y_0(x) - \frac{1}{x} Y_1(x).
\label{eqJYREC}
\eea

For $x \ll 1$:
\bea
J_0(x) &\approx& 1 \nonumber \\
Y_0(x) &\approx& \frac{2}{\pi} \left[\ln\left(\frac{x}{2}\right) + \gamma \right] \nonumber \\
J_1(x) &\approx& \frac{x}{2} \nonumber \\
Y_1(x) &\approx& -\frac{2}{\pi x} \nonumber \\
J_1'(x) &\approx& \frac{1}{2} \nonumber \\
Y_1'(x) &\approx& \frac{2}{\pi} \left[\frac{1}{x^2} + \ln\left(\frac{x}{2}\right) + \gamma \right]
\label{eqJYSMALL}
\eea
with $\gamma \approx 0.5772$. 

For $x \gg 1$
\beq
J_1(x) \approx \sqrt{\frac{2}{\pi x}} \cos\left(x-\frac{3\pi}{4}\right).
\label{eqJ1LARGE}
\eeq



\begin{thebibliography}{20}

\bibitem{Arkani-Hamed:1998rs}
N.~Arkani-Hamed, S.~Dimopoulos and G.~R.~Dvali,
Phys.\ Lett.\ B {\bf 429}, 263 (1998)
[arXiv:hep-ph/9803315];
I.~Antoniadis, N.~Arkani-Hamed, S.~Dimopoulos and G.~R.~Dvali,
Phys.\ Lett.\ B {\bf 436} (1998) 257
[arXiv:hep-ph/9804398].

\bibitem{Randall:1999ee}
L.~Randall and R.~Sundrum,
Phys.\ Rev.\ Lett.\  {\bf 83}, 3370 (1999)
[arXiv:hep-ph/9905221].

\bibitem{Goldberger:1999uk}
W.~D.~Goldberger and M.~B.~Wise,
Phys.\ Rev.\ Lett.\  {\bf 83}, 4922 (1999)
[arXiv:hep-ph/9907447].

\bibitem{Hall:TASI}
For an excellent review see: ``Aspects of Grand Unification'' by L.~Hall in TASI lectures in Elementary Particle Physics 1984. Edited by David N. Williams.

\bibitem{Dimopoulos:1981yj}
S.~Dimopoulos, S.~Raby and F.~Wilczek,
Phys.\ Rev.\ D {\bf 24}, 1681 (1981);
 W.~J.~Marciano and G.~Senjanovic,
Phys.\ Rev.\ D {\bf 25}, 3092 (1982).

\bibitem{Kaplan:1999ac}
D.~E.~Kaplan, G.~D.~Kribs and M.~Schmaltz,
Phys.\ Rev.\ D {\bf 62}, 035010 (2000)
[arXiv:hep-ph/9911293];
 Z.~Chacko, M.~A.~Luty, A.~E.~Nelson and E.~Ponton,
JHEP {\bf 0001}, 003 (2000)
[arXiv:hep-ph/9911323].

\bibitem{Barnett:1996hr}
R.~M.~Barnett {\it et al.}  [Particle Data Group Collaboration],
Phys.\ Rev.\ D {\bf 54}, 1 (1996).

\bibitem{Goldberger:1999wh}
W.~D.~Goldberger and M.~B.~Wise,
Phys.\ Rev.\ D {\bf 60}, 107505 (1999)
[arXiv:hep-ph/9907218].

\bibitem{Davoudiasl:1999tf}
H.~Davoudiasl, J.~L.~Hewett and T.~G.~Rizzo,
Phys.\ Lett.\ B {\bf 473}, 43 (2000)
[arXiv:hep-ph/9911262].

\bibitem{Gherghetta:2000qt}
T.~Gherghetta and A.~Pomarol,
Nucl.\ Phys.\ B {\bf 586}, 141 (2000)
[arXiv:hep-ph/0003129].

\bibitem{Pomarol:2000hp}
A.~Pomarol,
Phys.\ Rev.\ Lett.\  {\bf 85}, 4004 (2000)
[arXiv:hep-ph/0005293].

\bibitem{Randall:2001gc}
L.~Randall and M.~D.~Schwartz,
arXiv:hep-th/0108115.

\bibitem{Arkani-Hamed:2000ds}
N.~Arkani-Hamed, M.~Porrati and L.~Randall,
JHEP {\bf 0108}, 017 (2001)
[arXiv:hep-th/0012148].

\end{thebibliography}
\end{document}